\documentclass[journal]{IEEEtran}

\usepackage{amsfonts}
\usepackage{graphicx}
\usepackage{amsmath}
\usepackage{amssymb}
\begin{document}

		%\begin{frontmatter}

\title{Stable probability laws modeling random propagation times of waves crossing different media}

		\author{\IEEEauthorblockN{B. Lacaze} \\
		\IEEEauthorblockA{T\'{e}SA, 7 Boulevard de la Gare, 31500 Toulouse, France\\
		Email: bernard.lacaze@tesa.prd.fr}}

		%\author{B. Lacaze}

		%\address{T\'eSA, 7 Boulevard de la Gare, 31500 Toulouse, France\\
		%Email: bernard.lacaze@tesa.prd.fr}

		\maketitle

\begin{abstract}
In a communication scheme, there exist points at the transmitter and at the receiver where the wave is reduced to a finite set of functions of time which describe amplitudes and phases. For instance, the information is summarized in electrical cables which preceed or follow antennas. In many cases, a random propagation time is sufficient to explain changes induced by the medium. In this paper we study models based on stable probability laws which explain power spectra due to propagation of different kinds of waves in different media, for instance, acoustics in quiet or turbulent atmosphere, ultrasonics in liquids or tissues, or electromagnetic waves in free space or in cables. Physical examples show that a sub-class of probability laws appears in accordance with the causality property of linear filters.

		\textit{keywords:} random propagation time, stationary process, stable probability law, causality.
\end{abstract}

		%\begin{keyword}
		%		random propagation time, stationary process, stable probability law, causality.
		%\end{keyword}

		%\end{frontmatter}

\section{Introduction}

Acoustic waves propagate through shocks between molecules which move in the
medium. Consequently any (equivalent) time of propagation has a random
nature. The same property is true for electromagnetic waves which propagate
in a medium which is never totally empty and then which is submitted to
fluctuent fields. In both cases the result $\mathbf{U=}\left\{ U\left(
t\right) ,t\in \mathbb{R}\right\} $ of propagation of the wave $\mathbf{Z=}%
\left\{ Z\left( t\right) ,t\in \mathbb{R}\right\} $ can be modelled as%
\begin{equation}
U\left( t\right) =Z\left( t-A\left( t\right) \right)
\end{equation}%
where the random process $\mathbf{A=}\left\{ A\left( t\right) ,t\in \mathbb{R%
}\right\} $ is an equivalent propagation time.

This paper addresses stable probability laws modelling propagation times of
waves \cite{Loev}, \cite{Luka}, \cite{Zolo}, \cite{Nola}. The main effects
of propagation are weakenings, changes in phases and amplitudes, widenings
of power spectra and Doppler effects. I have shown that random propagation
times explained these transformations. For instance, in acoustics and
ultrasonics, where changes are ruled by the Beer-Lambert law, the (nearly)
Kramers-Kr\"{o}nig relations and selective attenuations \cite{Laca6}, \cite%
{Laca4}, \cite{Laca8}. Also in electromagnetics, where literature highlights
spectral widenings and/or Doppler shifts. Examples can be taken in radar
domain \cite{Laca9}, \cite{Laca11} or in optics for laser or natural light 
\cite{Laca7}, \cite{Laca10}. Propagation through coaxial cables shows a
particular form of weakenings and phase changes \cite{Laca12}. Let consider
an acoustic wave propagating in atmosphere or water. The path can be divided
into a sum of (approximately) independent pieces. The application of some
version of the central limit theorem leads to a Gaussian probability law for
the random propagation time \textbf{A}, the random behavior being due to
small and random changes of the refraction index. We show in section 3 that
this explains the weakening of pure tones in exp$\left[ -a\omega ^{2}\right] 
$ observed most of the time in acoustics ($\omega /2\pi $ is the frequency),
and many kinds of spectra \cite{Laca13}. But ultrasonics show more general
attenuations in exp$\left[ -a\omega ^{b}\right] ,0<b\leq 2,$ where $b$ is
often different of 2. Stable probability laws generalize the central limit
theorem and can be involved in this situation \cite{Loev}, \cite{Luka}, \cite%
{Zolo}. They can be asymmetric when $\alpha \neq 2$, and we will see that
this property allows to explain wave dispersion (the celerity depends on the
frequency).

The next section studies properties due to stationary random propagation
times. The Gaussian case and the more general case of stable laws are
explained in sections 3 and 4. Applications will be given in section 5 when
stable probability laws may be appealed. It is the case for acoustics,
ultrasonics and propagation through coaxial cables or in power cables, where experiments can be
done for wide enough frequency bands. These examples highlight a particular
sub-class of probability laws which is in accordance with the causality
property of linear filters. In the appendix, we explain the construction of
stationary stable processes which seem well matched to given examples.

\section{Random propagation times}

1) In what follows, we assume that $\mathbf{Z}$ (the real or complex random
process which propagates) is zero-mean and wide sense stationary with power
spectral density $s_{Z}\left( \omega \right) $ defined by%
\begin{equation}
\text{E}\left[ Z\left( t\right) Z^{\ast }\left( t-\tau \right) \right]
=\int_{-\infty }^{\infty }e^{i\omega \tau }s_{Z}\left( \omega \right)
d\omega .
\end{equation}%
E$\left[ ..\right] $ stands for the mathematical expectation (or ensemble
mean), the superscript $^{\ast }$ for the complex conjugate and $s_{Z}\left(
\omega \right) $ can be a mixing of regular functions and "Dirac functions". 
\textbf{A }(the random propagation time involved in $\left( 1\right) $) is a
random process independent of \textbf{Z }and such that the characteristic
functions%
\begin{equation}
\left\{ 
\begin{array}{c}
\text{E}\left[ e^{-i\omega A\left( t\right) }\right] =\psi \left( \omega
\right) \\ 
\text{E}\left[ e^{-i\omega (A\left( t\right) -A\left( t-\tau \right) )}%
\right] =\phi \left( \tau ,\omega \right)%
\end{array}%
\right.
\end{equation}%
are independent of $t.$ This defines a stationarity stronger than the usual
wide sense stationarity. The result of propagation \textbf{U }defined by $%
\left( 1\right) $ can be split in the orthogonal sum (see \cite{Laca1} and
the appendix)%
\begin{equation}
\begin{array}{c}
U\left( t\right) =G\left( t\right) +V\left( t\right) \\ 
\text{E}\left[ G\left( t\right) V^{\ast }\left( t-\tau \right) \right] =0%
\end{array}%
\end{equation}%
whatever $\tau \in \mathbb{R}$. The random process \textbf{G }is the output
of a linear invariant filter with input \textbf{Z }and complex gain $\psi
\left( \omega \right) .$ This means that%
\begin{equation}
\left\{ 
\begin{array}{c}
G\left( t\right) =\int_{-\infty }^{\infty }Z\left( t-u\right) f\left(
u\right) du \\ 
\psi \left( \omega \right) =\int_{-\infty }^{\infty }e^{-i\omega u}f\left(
u\right) du \\ 
s_{G}\left( \omega \right) =\left[ \left\vert \psi \right\vert ^{2}s_{Z}%
\right] \left( \omega \right)%
\end{array}%
\right.
\end{equation}%
for regular $f\left( u\right) $ (it is together the probability density of $%
-A\left( t\right) $ and the filter impulse response$).$ $s_{G}\left( \omega
\right) $ is the power spectral density of \textbf{G. }Moreover

\begin{equation}
\begin{array}{c}
K_{V}\left( \tau \right) =\text{E}\left[ V\left( t\right) V^{\ast }\left(
t-\tau \right) \right] = \\ 
\int_{-\infty }^{\infty }\left[ \phi \left( \tau ,\omega \right) -\left\vert
\psi \left( \omega \right) \right\vert ^{2}\right] e^{i\omega \tau
}s_{Z}\left( \omega \right) d\omega .%
\end{array}%
\end{equation}%
We may assume that 
\begin{equation*}
\lim_{\tau \rightarrow \infty }\phi \left( \tau ,\omega \right) =\left\vert
\psi \left( \omega \right) \right\vert ^{2}
\end{equation*}%
with a fast enough convergence. This means that $A\left( t\right) $ and $%
A\left( t-\tau \right) $ become independent when $\tau $ is large enough. In
this case, \textbf{V }will have a regular power spectral density $%
s_{V}\left( \omega \right) $ defined by%
\begin{equation}
K_{V}\left( \tau \right) =\int_{-\infty }^{\infty }e^{i\omega \tau
}s_{V}\left( \omega \right) d\omega .
\end{equation}

2) Let assume that devices transmit a pure spectral line at the frequency $%
\omega _{0}/2\pi $ 
\begin{equation*}
Z\left( t\right) =e^{i\omega _{0}t}.
\end{equation*}%
The fact that the mean depends on $t$ has no importance. The received wave
verifies, from $\left( 5\right) ,\left( 6\right) $%
\begin{equation}
\left\{ 
\begin{array}{c}
G\left( t\right) =e^{i\omega _{0}t}\psi \left( \omega _{0}\right) \\ 
2\pi s_{V}\left( \omega \right) = \\ 
\int_{-\infty }^{\infty }\left[ \phi \left( \tau ,\omega _{0}\right)
-\left\vert \psi \left( \omega _{0}\right) \right\vert ^{2}\right]
e^{i(\omega _{0}-\omega )\tau }d\tau .%
\end{array}%
\right.
\end{equation}%
The result of the propagation is the sum of a pure wave \textbf{G} (which
ends to a DC part after an amplitude demodulation of shift $\omega _{0}$)
and a band process \textbf{V} (the AC part). The powers of the components
are equal to $\left\vert \psi \left( \omega _{0}\right) \right\vert ^{2}$
(for the DC one) and 
\begin{equation*}
K_{V}\left( 0\right) =1-\left\vert \psi \left( \omega _{0}\right)
\right\vert ^{2}
\end{equation*}%
(for the AC one). The sum of powers is equal to 1, i.e. the propagation is
lossless.

The DC/AC ratio $r$ is the ratio of the respective powers \cite{Bill}%
\begin{equation}
r=\frac{\left\vert \psi \left( \omega _{0}\right) \right\vert ^{2}}{%
1-\left\vert \psi \left( \omega _{0}\right) \right\vert ^{2}}.
\end{equation}%
Actually, devices of reception take into account a limited band around the
frequency $\omega _{0}.$ As a consequence, it happens that only the
monochromatic part is viewed in more or less important noise which is the
sum of a surrounding medium noise and a little part of the companion process 
$\mathbf{V}$ defined by $\left( 4\right) ,\left( 5\right) $ and $\left(
6\right) \mathbf{.}$ It is often the case in acoustics \cite{Laca6} and in
ultrasonics \cite{Laca8} or in propagation in coaxial cables \cite{Laca12}.
Conversely the spectral line can disappear. For instance for electromagnetic
waves in free space. Light coming from stars shows lines of hydrogen $%
H_{\alpha },H_{\beta }...$ and of other elements with various widths \cite%
{Laca10}. We know that laser emissions are broadened when propagating in
space \cite{Ridl}. Moreover, in the computation of power spectra, the
autocorrelation function is cancelled for missing data (apodization). This
leads to a widening of pure spectral lines which can disappear in the
surrounding noise \cite{Laca13}.

\section{The Gaussian case}

1) Let assume that \textbf{A} is a Gaussian process. If the propagation time
can be considered as a sum of approximately independent components (for
instance successive thicknesses in the atmosphere), its mean, its variance
and its covariance are proportional to the propagation distance $l.$ We
assume that these quantities depend only on the crossed medium (and not on
the wave properties). Then, the characteristic functions of $\left( 3\right) 
$ are defined by%
\begin{equation}
\left\{ 
\begin{array}{c}
\psi \left( \omega \right) =\exp \left[ -iml\omega -l\left( \sigma \omega
\right) ^{2}/2\right] \\ 
\phi \left( \tau ,\omega \right) =\exp \left[ -l\left( \sigma \omega \right)
^{2}\left( 1-\rho \left( \tau \right) \right) \right] \\ 
l\sigma ^{2}\rho \left( \tau \right) =\text{E}\left[ A\left( t\right)
A\left( t-\tau \right) \right]%
\end{array}%
\right.
\end{equation}%
where $m$ is the mean, $\sigma $ the standard deviation and $\rho \left(
\tau \right) $ the (normalized) autocorrelation function for the unit
distance. From $\left( 8\right) $, we have%
\begin{equation}
G\left( t\right) =e^{i\omega _{0}(t-ml)-l\left( \sigma \omega _{0}\right)
^{2}/2}.
\end{equation}%
$ml$ is the transit time of the pure wave on the distance $l.$ The celerity
of the wave \textbf{G} is equal to $m^{-1}$ and does not depend on the
frequency$.$ Equivalently the refraction index is constant (the medium is
not dispersive). The dependence on $l$ of the amplitude and the phase of $%
G\left( t\right) $ is a proof that we are in the Beer-Lambert law context.\
Beside the monochromatic wave \textbf{G} appears the process \textbf{V }with%
\textbf{\ }power\textbf{\ }spectral density defined by (from $\left(
8)\right) $%
\begin{equation}
\begin{array}{c}
2\pi s_{V}\left( \omega \right) = \\ 
\int_{-\infty }^{\infty }e^{-i(\omega -\omega _{0})\tau -l\left( \sigma
\omega _{0}\right) ^{2}}\left[ e^{l\left( \sigma \omega _{0}\right) ^{2}\rho
\left( \tau \right) }-1\right] d\tau .%
\end{array}%
\end{equation}%
It is possible to choose $\rho \left( \tau \right) $ in sort that $%
s_{V}\left( \omega \right) $ takes very small values in the frequency bands
viewed by devices. Then, \textbf{V} will be neglected and/or plugged in
other surrounding noises.

2) Let assume that the behavior of $\rho \left( \tau \right) $ at the origin
point verifies%
\begin{equation}
\rho \left( \tau \right) =1-\left\vert \frac{\tau }{\tau _{0}}\right\vert
^{b}+o\left( \left\vert \frac{\tau }{\tau _{0}}\right\vert ^{b}\right)
\end{equation}%
with $0<b\leq 2.$ $\left( 13\right) $ rules the behavior of $\rho \left(
\tau \right) $ nearby the point $\tau =0.$ $\rho \left( \tau \right) $ takes
into account the celerity of variations of the propagation time \textbf{A }%
by length unit\textbf{. }Practically, from $\left( 12\right) $, this means
that we have 
\begin{equation}
\begin{array}{c}
s_{V}\left( \omega _{0}+\omega \frac{\mu ^{1/b}}{\tau _{0}}\right) \cong \\ 
\frac{\tau _{0}\mu ^{-1/b}}{2\pi }\int_{-\infty }^{\infty }e^{-i\omega
u-\left\vert u\right\vert ^{b}}du \\ 
\mu =l\left( \omega _{0}\sigma \right) ^{2}%
\end{array}%
\end{equation}%
when $\mu $ is sufficiently large. In the same time, the bandwidth of 
\textbf{V }increases when $\tau _{0}$ decreases$.$ $\tau _{0}$ has the same
meaning as a "decorrelation time".

3) To a better understanding, let assume that $\omega _{0}/2\pi =10^{6}$
(for instance in ultrasonics), $l=1$ and%
\begin{equation*}
\omega _{0}\sigma =1,\rho \left( \tau \right) =\left\{ 
\begin{array}{c}
1-\left\vert \tau /\tau _{0}\right\vert ,\left\vert \frac{\tau }{\tau _{0}}%
\right\vert <1 \\ 
0,\left\vert \tau /\tau _{0}\right\vert \geq 1 \\ 
\tau _{0}=10^{-9}\text{.}%
\end{array}%
\right.
\end{equation*}%
We have respectively $P_{G}=e^{-1}$ (it is the power of \textbf{G, }the
height of the residual spectral line\textbf{),} $P_{G}+P_{V}=1$ ($P_{V}$ is
the power of \textbf{V). }$s_{V}\left( \omega \right) $ is almost constant
in the interval $\left( \omega _{0}-10^{8},\omega _{0}+10^{8}\right) $ with
value less than 10$^{-9}$. Clearly, the process \textbf{V }cannot be
distinguished by devices matched to the extraction of the line \textbf{G }%
(i.e centered around the frequency $\omega _{0}/2\pi )$\textbf{. }

Conversely, when $\omega _{0}/2\pi =10^{14},l=1$ (for instance in infrared),
let assume that 
\begin{equation*}
\omega _{0}\sigma =4,\rho \left( \tau \right) =\left\{ 
\begin{array}{c}
1-\left\vert \tau /\tau _{0}\right\vert ,\left\vert \tau \right\vert <\tau
_{0} \\ 
0,\left\vert \tau \right\vert \geq \tau _{0}.%
\end{array}%
\right.
\end{equation*}%
We obtain $P_{G}\simeq 0,P_{V}\simeq 1$. Furthermore, provided that $\tau
_{0}$ is not too large 
\begin{equation*}
\pi s_{V}\left( \omega +\omega _{0}\right) \simeq \frac{\theta }{\theta
^{2}+\omega ^{2}},\theta =\frac{16}{\tau _{0}}.
\end{equation*}%
$\omega _{0}$ is the gravity center of $s_{V}\left( \omega \right) ,$ and $4/%
\sqrt{\tau _{0}}$ is a measure of the bandwidth. If $l=2$ and not $l=1,$ we
obtain $\theta =32/\tau _{0}.$ This means that increases of $l$ (like
decreases of $\tau _{0})$ lead to a flattening (a widening) of $s_{V}\left(
\omega \right) .$ The Lorentzian behavior of $s_{V}\left( \omega \right) $
is related to the value $b=1$ in $\left( 13\right) .$ $(14)$ defines the
links between the value of $b$ and the shape of $s_{V}\left( \omega \right)
. $ For instance $b=2$ leads to a Gaussian spectrum.

\section{Stable probability laws}

1) The probability law of $A\left( t\right) $ is stable when the
characteristic function (c.f) $\psi \left( \omega \right) =$E$\left[
e^{-i\omega A\left( t\right) }\right] $ defined in $\left( 3\right) $ has
the shape \cite{Loev}, \cite{Luka}, \cite{Zolo}%
\begin{equation}
\begin{array}{c}
\ln \psi \left( \omega \right) = \\ 
-im\omega -a\left\vert \omega \right\vert ^{\alpha }\left( 1+i\beta \frac{%
\omega }{\left\vert \omega \right\vert }h\left( \left\vert \omega
\right\vert ,\alpha \right) \right)%
\end{array}%
\end{equation}%
with real parameters $\left( -m,a,\alpha ,\beta \right) $ and $a>0,0<\alpha
\leq 2,\left\vert \beta \right\vert \leq 1.$ Moreover%
\begin{equation}
h\left( \left\vert \omega \right\vert ,\alpha \right) =\left\{ 
\begin{array}{c}
\tan \frac{\pi \alpha }{2},\alpha \neq 1 \\ 
\frac{2}{\pi }\ln \left\vert \omega \right\vert ,\alpha =1.%
\end{array}%
\right.
\end{equation}%
The corresponding random variables have probability densities with analytic
closed form in three cases, $\alpha =2$ (Gaussian), $\alpha =1,\beta =0$
(Cauchy/ Lorentz) and $\alpha =1/2,\beta =\pm 1$ (Levy). In other cases,
series developments can be used \cite{Luka}, \cite{Zolo}. The laws are
always unimodal and they are one-sided only when $0<\alpha <1,\beta =\pm 1.$
Formula $\left( 15\right) $ is matched to an unit propagation length. For a
length $l,$ we can replace $A$ by $A_{l},$ $m$ by $ml$ and $a$ by $al.$ This
means that the r.v $A_{l}\left( t\right) ,$ the propagation time on a length 
$l$ is the sum of $l$ r.v which have the same probability law that $A\left(
t\right) $ (for any integer $l).$

In the notations of \cite{Luka}, $\psi \left( \omega \right) $ is the c.f of
the r.v $-A\left( t\right) .$ Then, the opposite $A\left( t\right) $ follows
a stable law of parameters $\left( m,a,\alpha ,-\beta \right) .$ We will see
that the value of $\beta $ has a great importance. When $\beta =1,0<\alpha
<1,m=0,$ it is proved that$\ $the r.v $A\left( t\right) $ is such that 
\begin{equation*}
\text{Pr}\left[ A\left( t\right) >0\right] =1.
\end{equation*}%
Equivalently, the filter of complex gain $\psi \left( \omega \right) $ is
causal, because the impulse response of this filter is the probability
density of $A\left( t\right) $ (look at $\left( 3\right) $ and $\left(
6\right) ).$

For $1<\alpha <2,$ the r.v $A\left( t\right) $ is never one-sided.
Nevertheless, the value $\beta =1$ provides the most asymmetric possible
probability density, so that this value is the best choice with respect to
the causality property. When $\alpha =1,$ we have to take the value $\beta
=-1,$ which gives the same property. For $\alpha =2,$ the value of $\beta $
is indifferent.

2) The choice of this class of probability laws will be justified in the
following section by practical examples. Experiments often show that the
r.v. $A\left( t\right) $ has to be Gaussian for each $t$ (the particular
case $\alpha =2)$. In this situation it is natural to assume that the entire
process \textbf{A }is Gaussian which implies that the margins $A\left(
t\right) $ and the linear combinations $A\left( t\right) -A\left( t-\tau
\right) $ are Gaussian. It is not a great bet though the knowledge of
margins does not define entirely the underlying probability laws most of the
time. This kind of problem is complex \cite{Nels}.

The propagation of ultrasonic waves involves stable laws more general than
the Gaussian one, see section 5.2 and \cite{Laca4}, \cite{Laca8}. In
appendix, we prove that it is possible to construct processes \textbf{A }%
such that both $A\left( t\right) $ and $A\left( t\right) -A\left( t-\tau
\right) $ follow stable laws with the same exponent $\alpha ,$ whatever its
value in $]0,2].$ This means that the Gaussian model can be naturally
generalized. To take $\alpha $ different of $2$ in $\left( 15\right) $ does
not really limit the possibilities for spectral densities of \textbf{V }%
(defined in section 2). In this expanded context it is possible to model
propagations with mixed results (neither \textbf{G }nor\textbf{\ V }is
negligible \cite{Laca14}).

3) We have seen in section 2 that $\psi \left( \omega \right) $ can be
viewed as the complex gain of a linear invariant filter (which provides the
process \textbf{G}). $\psi \left( \omega _{0}\right) \exp \left[ i\omega
_{0}t\right] $ is the answer of some device to the pure tone $\exp \left[
i\omega _{0}t\right] .$ In the Gaussian framework which is described by $%
\left( 10\right) $, the wave amplitude is weakened by $\exp \left[ -l\omega
_{0}^{2}\sigma ^{2}/2\right] $ and delayed by the constant $ml.$ $c=1/m$ is
the celerity of the wave in the medium and does not depend on $\omega _{0}$.
More generally, when $\psi \left( \omega \right) $ verifies $\left(
15\right) ,$ the wave weakening is equal to%
\begin{equation*}
\exp \left[ -a\left\vert \omega _{0}\right\vert ^{\alpha }\right] .
\end{equation*}%
The propagation time of the monochromatic part \textbf{G} is no longer a
constant (apart when $\alpha =2)$ but it becomes the quantity ($\omega
_{0}>0)$%
\begin{equation*}
m+a\beta \omega _{0}^{\alpha -1}\left\{ 
\begin{array}{c}
\tan \left( \pi \alpha /2\right) ,\alpha \neq 1 \\ 
\frac{2}{\pi }\ln \omega _{0},\alpha =1.%
\end{array}%
\right. 
\end{equation*}%
Except for $\alpha =2$ (the Gaussian case), this means that we deal with a
dispersive medium in which the celerity is a function of the frequency.
Experiments described below prove that the parameter $\beta $ in $\left(
15\right) $ has to take one of particular values $1$ or $-1.$ It is of huge
importance for the model validity because these values are closely linked to
the notion of causality (i.e the consequence of an action appears only after
the action).

\begin{figure}[htb]
		\centering
		\includegraphics[width=8.5cm]{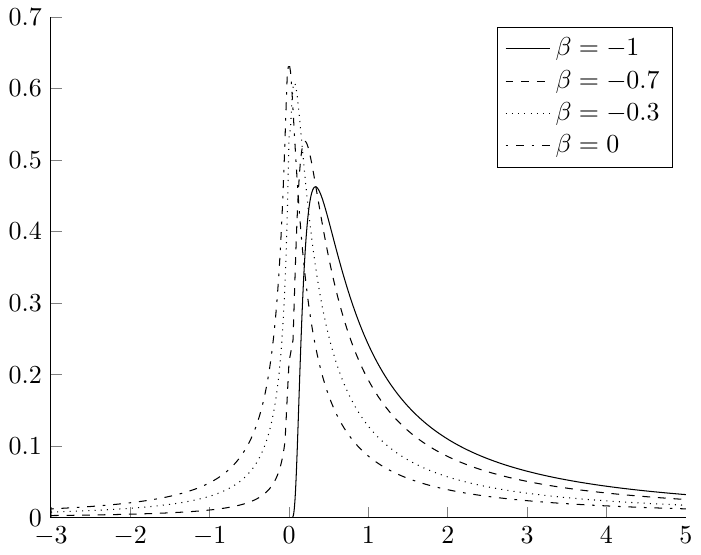}
		\caption{Probability density of $A(t)$ for $\alpha=\frac{1}{2}$, $a=1$ as a function of $\beta$}
		\label{fig_1}
\end{figure}

\begin{figure}[htb]
		\centering
		\includegraphics[width=8.5cm]{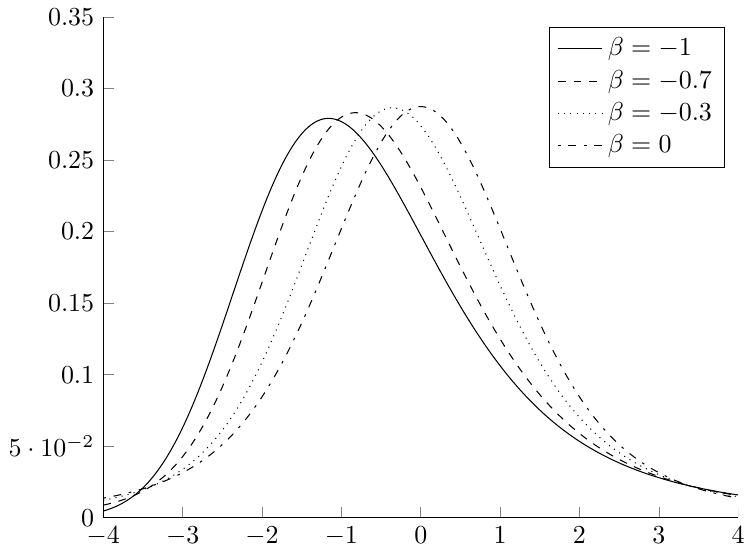}
		\caption{Probability density of $A(t)$ for $\alpha=\frac{3}{2}$, $a=1$ as a function of $\beta$}
		\label{fig_2}
\end{figure}

\begin{figure}[htb]
		\centering
		\includegraphics[width=8.5cm]{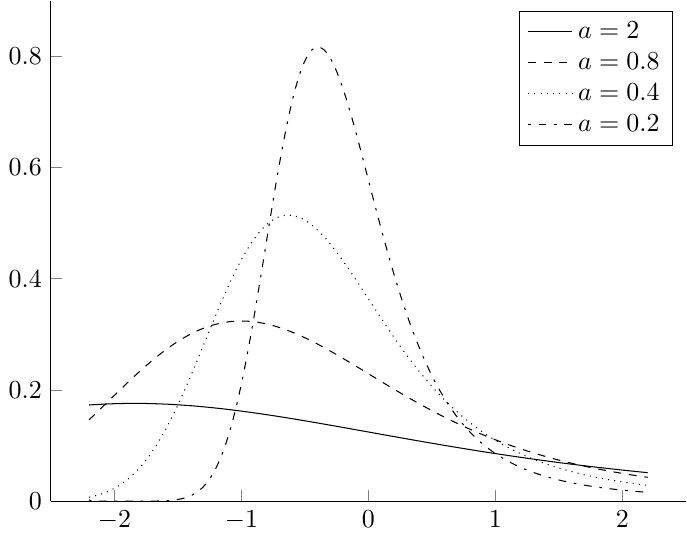}
		\caption{Probability density of $A(t)$ for $\alpha=\frac{3}{2}$, $\beta=-1$ as a function of $a$}
		\label{fig_3}
\end{figure}

4) Figures \ref{fig_1}, \ref{fig_2}, \ref{fig_3} illustrate stable laws properties linked to probability
densities (which are always regular and unimodal). Figure 1 depicts the case
(0,1,1/2,$\beta ),$ for some values of $\beta .$ The value $\beta =-1$
provides the only one-sided probability density (which leads to causality).
In Figure 2, the case (0,1,3/2,$\beta )$ is studied. The value $\beta =-1$
does not provide a one-sided probability density, but it is the best bet for
approaching the causality property. Figure 3 shows the case (0,$a,3/2,-1).$
We see that the "width" of the probability density decreases in the same
time that $a.$

\section{Examples}

\subsection{Acoustics}

No propagation phenomenon was better studied than acoustic waves through
atmosphere and sea water (mainly for military aims in the latter case). In
frequency bands where physical conditions in the medium are invariant, the
complex amplitude $F_{l}\left( \omega \right) $ of the monochromatic wave $%
e^{i\omega t}$ measured at the distance $l$ is in the form ($\omega /2\pi $
is the frequency)%
\begin{equation}
F_{l}\left( \omega \right) =e^{-l[a\omega ^{2}+i\omega /c]}
\end{equation}%
where $a$ depends on the temperature, viscosity, composition.... The
celerity $c$ of the wave has the same property \cite{Kins}, \cite{Uric}. $%
\left( 17\right) $ is true for frequencies going up to dozens of MHz and
whatever the path length $l$ up to kilometres if the medium is subjected to
constant conditions. These conditions are able to change because some kinds
of molecules in atmosphere (H$_{2}$O, CO$_{2}..)$ or in water (MgSO$_{4},$
B(OH)$_{3}..)$ interact at characteristic frequencies. $a$ is constant in
frequency bands which depend on the kind of molecules and on their
concentrations. In frequency bands where $a$ and $c$ can be considered as
constant, $\left( 17\right) $ shows that the medium on a length $l$ reacts
like a linear invariant filter of complex gain $F_{l}\left( \omega \right) .$
Consequently, the Gaussian model of propagation studied in section 3 is
available. The random propagation time \textbf{A }is a Gaussian process with 
$\psi \left( \omega \right) =F_{l}\left( \omega \right) .$ The received wave
is the component \textbf{G. }The component \textbf{V }has likely a flat
power spectrum which makes it invisible among ambient noises, because
measurements devices are matched to the transmitted frequency. Let note that
literature is very detailed for variations of $a$ with respect to frequency,
temperature and other parameters. But I do not know references about
variations of $c$ versus the frequency. This means that we have to admit
that the medium is not dispersive in this occurence.

\subsection{Ultrasonics}

Ultrasonic waves crossing various materials on small distances (for instance
for echography) lead to more general formulas \cite{Laca8}. In this case, we
often measure complex amplitudes $F_{l}\left( \omega \right) $ (for
monochromatic transmitted waves at the frequency $\omega /2\pi $ on a length 
$l$)%
\begin{equation}
\begin{array}{c}
F_{l}\left( \omega \right) =e^{-i\omega (l/c)-al\left\vert \omega
\right\vert ^{\alpha }\left( 1+i(\text{sgn}\omega )\tan \left( \pi \alpha
/2\right) \right) } \\ 
0<\alpha \leq 2,\alpha \neq 1.%
\end{array}%
\end{equation}%
or, for $\alpha =1$%
\begin{equation*}
F_{l}\left( \omega \right) =e^{-i\omega (l/c)-al\left\vert \omega
\right\vert \left( 1-\frac{2}{\pi }i(\text{sgn}\omega )\ln \left\vert \omega
\right\vert \right) }
\end{equation*}%
We recognize particular cases of $\left( 15\right) $ about stable
probability laws with $\beta =\pm 1$.

This model is available with $\beta =1$ for very different media like castor
oil ($\alpha =1.66)$, polyethylene ($\alpha =1.1)$, rubber ($\alpha =1.38)$
and also brass tube ($\alpha =0.5$ but for acoustic low frequencies) \cite%
{Szab}, \cite{He}, \cite{Maso}. The case $\alpha =1$ (with $\beta =-1)$ is
encountered for instance when modeling evaporated milk. The value $\beta
=\pm 1$ is often well-fitted and has a particularly physical meaning. When $%
\alpha <1,$ the probability density of $F_{l}\left( \omega \right) $ is
one-sided (it is false for other values of $\beta )$. Equivalently, the
linear invariant filter of complex gain $F_{l}\left( \omega \right) $\textbf{%
\ }is causal (the result \textbf{G }is viewed by devices). When $\alpha \geq
1,$ these values of $\beta $ minimize a tail of distribution or equivalently
they are the most favourable with respect to the causality property. Of
course it is a huge argument for physicists. Values of $\beta $ different of 
$\pm 1$ are also encountered for example for egg yolk ($\alpha =1.54,\beta
=0.85)$ and lucite ($\alpha =0.625,\beta =0.71)$ \cite{Laca8}$.$ But
measurements of the celerity in the medium which rules the value of $\beta $
are difficult and can be less accurate than measurements of $\alpha $ which
rules the attenuation variations$.$ Like in the acoustic case, the second
component (noted \textbf{V} in section 2) is not detected by devices. This
means that the \textbf{V}-power spectrum is very flat and spread with
respect to the frequencial window of reception devices.

\subsection{Coaxial cables}

Propagation in coaxial cables can be compared to propagation of acoustic or
ultrasonic waves (Drude model). In the latter cases, the propagation is due
to interactions between molecules and in the first case, to interactions
between electrons of conduction and the medium. From a classical theory and
for an infinite line, a length $l$ of cable is equivalent to a linear
invariant filter of complex gain%
\begin{equation*}
F_{l}\left( \omega \right) =e^{-l\sqrt{\left( R+iL\omega \right) \left(
G+iC\omega \right) }}
\end{equation*}%
where $R,L,C,G$ are the series resistance, the series inductance, the shunt
capacitance and the shunt conductance by length unit. Actually, $R$ depends
on $\omega $ due to the "skin effect", $G$ is small, which leads to the
formula%
\begin{equation}
F_{l}\left( \omega \right) =e^{-l\left( im\omega +a\sqrt{\left\vert \omega
\right\vert }\left( 1+i\text{sgn}\omega \right) \right) }.
\end{equation}%
This formula is for example available for the Belden 8281 cable up to 1GHz 
\cite{Liu}, \cite{Laca12}. Then we see that $-A\left( t\right) $ follows a
stable probability law of parameters ( see section 4) 
\begin{equation*}
\left( -ml,al,1/2,1\right) .
\end{equation*}%
The parameter $1$/$m$ (the wave celerity) is close to 2/3 the light celerity
in the vacuum, $a$ depends on the manufacturing process. The value $\alpha
=1/2$ is not encountered in ultrasonics (at my knowledge we have $\alpha
\geq 1$ in this domain) but we find this value in \cite{Maso} (propagation
in brass tubes at low frequency). We have explained in the section above
that the value $\beta =1$ has a particular sense. It means that the
propagation on a distance $l$ is equivalent to a causal linear invariant
filter with complex gain $F_{l}\left( \omega \right) $. The causality
property is verified only for this value of $\beta .$ As usual, the output
of this filter is identified to the process \textbf{G }defined in section 2.
The process \textbf{V} summarizes losses which are not taken into account by
devices.

\begin{figure}[htb]
\centering
\includegraphics[width=8.3cm]{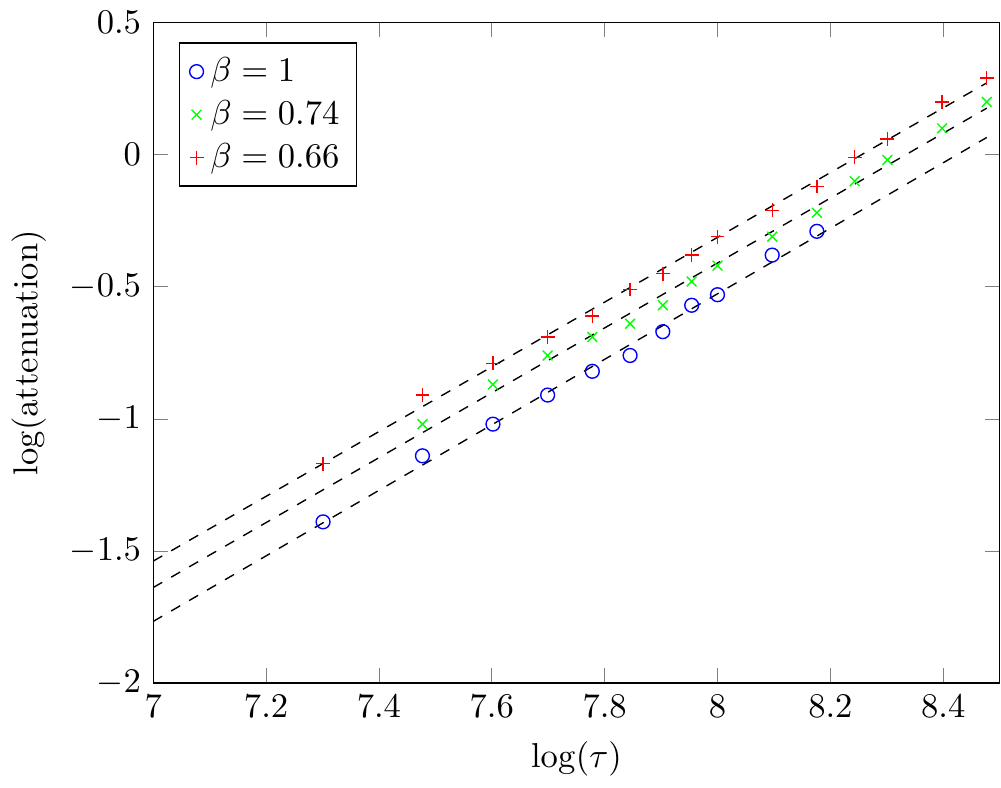}
\caption{Log(attenuation) plot of XLPE cable from measurements in \cite{Papa}}
\label{fig_4}
\end{figure}

The stable probability law model is available for other conductors like
XLPE cables for energy transmission. In figure 1 of \cite{Ouss}, attenuation
versus frequency is well-fitted by $y=ax^{\alpha }$ (in dB) with $\alpha
=1.13$ up to 50MHz for a cable in a good state of repair$.$ Unfortunately,
the authors do not give curves about the celerity, and it is not possible to
verify the shape of the imaginary part in $\left( 15\right) $ (and then the
value of the parameter $\beta ).$ Figure \ref{fig_4} in this paper shows the fit of
attenuation for data in \cite{Papa} and three species of cables in good
state. We find $\alpha =1.23-1.24.$ Moreover the study of data giving the
variation of the wave celerity leads to the values 
\begin{equation*}
\beta =1-0.74-0.66.
\end{equation*}%
Recall that the causality of filters corresponds to the value $\beta =1.$

\subsection{The Mintzer's law}

Propagation of acoustics in water or atmosphere is well explained by
Gaussian propagation times (see section 5.1). The parameters are defined by
the macroscospic characteristics of the medium, like temperature, viscosity,
composition.... About the propagation of the monochromatic wave $e^{i\omega
_{0}t}$ on the distance $l,$ the amplitude is constant and equal to $%
e^{-al\omega _{0}^{2}}$ (see section 5.1). In the case of a turbulent
medium, for instance induced by a heated grid \cite{Andr}, we may consider
that the weakening parameter $a$ is a slowly time-variant quantity $B\left(
t\right) $, which has a random behavior linked to the strength of
turbulences. Consequently, the wave amplitude $C\left( t\right) $ at the
distance $l$ becomes the random quantity%
\begin{equation}
C\left( t\right) =\exp \left[ -B\left( t\right) l\omega _{0}^{2}\right] .
\end{equation}%
Classically, the "coefficient of variation" $\mu $ of the r.v (random
variable) $C\left( t\right) $ is defined by \cite{Bill} 
\begin{equation*}
\mu =\frac{\sqrt{\text{Var}C\left( t\right) }}{\text{E}\left[ C\left(
t\right) \right] }.
\end{equation*}%
The Mintzer's law asserts that $\mu $ \ varies like $\omega _{0}\sqrt{l}$ 
\cite{Mint}. We ask to find probability laws for $B\left( t\right) $ which
are able to explain such a result (we assume that the process $\mathbf{B}$
is stationary).

Let assume the existence of the moment-generating function%
\begin{equation*}
\xi \left( s\right) =\text{E}\left[ e^{-sB\left( t\right) }\right] ,s\geq 0.
\end{equation*}%
It is a strong hypothesis when the probability laws are not one-sided \cite%
{Zolo}. We have the equality%
\begin{equation}
\mu =\frac{\sqrt{\xi \left( 2l\omega _{0}^{2}\right) -\xi ^{2}\left( l\omega
_{0}^{2}\right) }}{\xi \left( l\omega _{0}^{2}\right) }
\end{equation}%
When $B\left( t\right) $ follows a stable law with parameters $\left(
m^{\prime },\theta ,1,1\right) ,$ we know that $\xi \left( s\right) $ exists
and that \cite{Zolo}%
\begin{equation}
\xi \left( s\right) =\exp \left[ \theta s\ln s-m^{\prime }s\right] .
\end{equation}%
In this case, $\left( 21\right) $ becomes%
\begin{equation*}
\mu =\sqrt{\exp \left[ 2\theta l\omega _{0}^{2}\ln 2\right] -1}.
\end{equation*}%
For small $\mu $ corresponding to small $l\omega _{0}^{2},$ we have%
\begin{equation}
\mu \sim \sqrt{2\theta \ln 2}\omega _{0}\sqrt{l}
\end{equation}%
in agreement with the Mintzer's law. Experimental values found in literature
(see \cite{Laca6}) justify the limited development. For stable laws with
parameters $\left( m,\theta ,\alpha ,-1\right) ,\alpha \neq 1,$ $\xi \left(
s\right) $ exists (and not in other cases, except the preceeding one \cite%
{Zolo}), and we obtain%
\begin{equation*}
\mu \sim \sqrt{\theta \left\vert 2-2^{\alpha }\right\vert }\left( l\omega
_{0}^{2}\right) ^{\alpha /2}
\end{equation*}%
which returns to $\left( 23\right) $ when $\alpha \rightarrow 1.$ In \cite%
{Laca6}, we show that the half-Cauchy law, which is not stable, leads to the
Mintzer's property. It is a law with discontinuous probability density, but
which has the advantage to be one-sided (it is not true for $\left(
22\right) )$. Let note that physicists study moments of the log-amplitude.
In the situations which are developed in this section, these quantities do
not exist.

\subsection{Backscatter from trees}

\begin{figure}[htb]
\centering
\includegraphics[width=8.5cm]{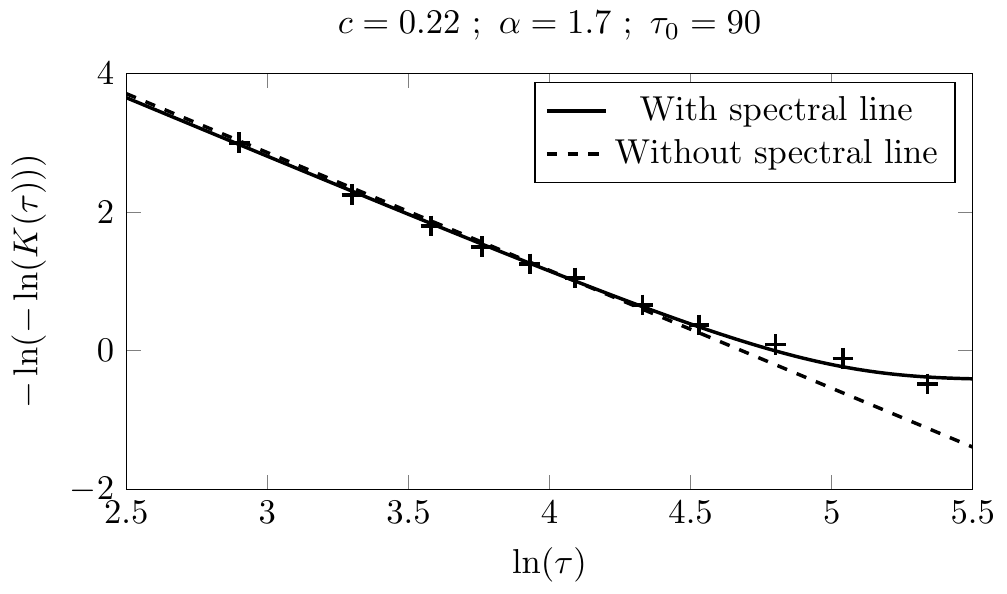}
\caption{Autocorrelation plot of apple tree at a windspeed of $3.8$ m/s}
\label{fig_5}
\end{figure}

\begin{figure}[htb]
\centering
\includegraphics[width=8.5cm]{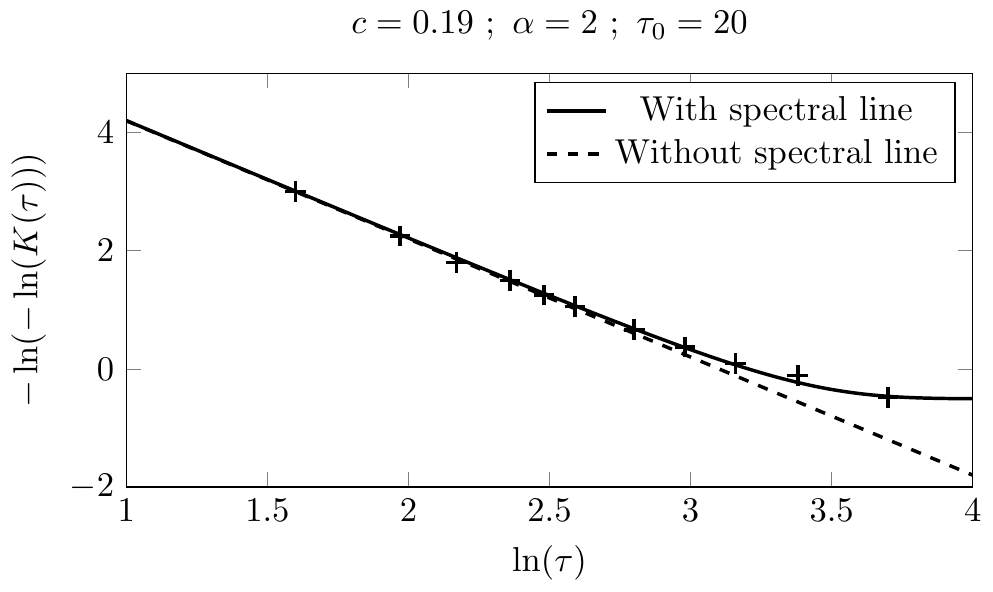}
\caption{Autocorrelation plot of white mulberry at a windspeed of $7.9$ m/s}
\label{fig_6}
\end{figure}

In paper \cite{Nara}, a radar at 8GHz interacts with crowns of trees. Each
experiment addresses a particular tree among six species, and at two
different wind celerities. Measurements of autocorrelation functions $%
K\left( \tau \right) $ of the backscatter are compared with the so-called
Wong model (a mixture of Gaussians) \cite{Laca15}. If we consider the curves
up to the "decorrelation time" $\tau _{0}$ (defined by $K\left( \tau
_{0}\right) =1/e),$ they are very well fitted by (after demodulation in
baseband) 
\begin{equation*}
K\left( \tau \right) =\exp \left[ -\left( \tau /\tau _{0}\right) ^{\alpha }%
\right] 
\end{equation*}%
where 1.65$\leq \alpha \leq 2,$ following the specy of tree and the wind
speed \cite{Laca9}. We know that the accuracy of measurements decreases with 
$\tau ,$ due to apodization and variations of the frequency. It is the
reason which leads to neglect measurements above the decorrelation time, and
in the same time to admit that lim$_{\tau \rightarrow \infty }K\left( \tau
\right) =0.$ Nevertheless, all data of \cite{Nara} drawn in the set of
coordinates 
\begin{equation*}
\left( x=\ln \tau ,y=-\ln \left[ -\ln K\left( \tau \right) \right] \right) 
\end{equation*}
are above the line%
\begin{equation*}
y=\alpha x-\alpha \ln \tau _{0}
\end{equation*}%
which cannot be due only to inaccuracies. Taking into account data above $%
\tau _{0}$ leads to the model%
\begin{equation*}
K\left( \tau \right) =c+\left( 1-c\right) \exp \left[ -\left( \tau /\tau
_{0}\right) ^{\alpha }\right] .
\end{equation*}%
We see that a specular component appears with a power equal to $c\ $and a
DC/AC\ ratio $r$ equal to (see$\left( 9\right) )$%
\begin{equation*}
r=c/\left( 1-c\right) .
\end{equation*}%
Figures \ref{fig_5} and \ref{fig_6} from \cite{Laca13}\ illustrate the fits for apple tree and white mulberry. Gaussian
propagation times explain these results. Neither the $\mathbf{G}$ part (here
the DC component) nor the \textbf{V }part (the AC component)\textbf{\ }%
disappears. 

\subsection{Electromagnetic waves in free space}

Generally, propagation of electromagnetic waves in free space or atmosphere
leads to weakenings, spectral widenings and Doppler shifts. The
disappearance of spectral lines is easily explained by Gaussian random
propagation times. The weakening of the line $\exp \left[ i\omega _{0}t%
\right] $ is measured by the quantity exp$\left[ -\omega _{0}^{2}\sigma
^{2}/2\right] $ where the standard deviation $\sigma =\sqrt{\text{Var}%
A\left( t\right) }$ characterizes the amplitude of time variations of the
wave trajectory. For a He/Ne laser at 0.633$\mu $m (in the visible) on a
propagation distance of 15m, $\omega _{0}\sigma =10$ for $\sigma =3.10^{-14}$%
s. This value corresponds to 10$^{-6}$ times the mean propagation time,
which is likely much smaller than uncertainties about the propagation
length. Clearly, the line disappears and is totally replaced by a continuous
spectral component. In this situation, the characteristic function $\psi
\left( \omega \right) $ cannot be measured. Even when a part of the line can
be highlighted \cite{Laca13}, the transmitters (radar, laser,...) work
generally at only one frequency, which does not allow to measure functions
of the frequency. Nevertheless, the classical central limit theorem works
for acoustics and ultrasonics in free space ($\omega _{0}$ has not the same
order of magnitude, it remains a pure spectral line, see sections 5.1 and
5.2), and, because the lack of data, the Gaussian model seems the best bet
about electromagnetic propagation in free space. In this context, the
behavior close to the origin of $\rho \left( \tau \right) $ defined in $%
\left( 13\right) ,$ allows to fit spectral densities in the form%
\begin{equation*}
K_{V}\left( \tau \right) =a\exp \left[ -b\tau ^{\alpha }\right]
\end{equation*}%
where $0<\alpha \leq 2$ \cite{Laca13}. Moreover, Doppler shifts appear when
moving targets and /or continuous spectra of transmitters \cite{Laca9}, \cite%
{Laca10}, \cite{Laca11}.

\section{Remarks}

1) We can admit that, when the model of stable laws is available, the value
of $\beta $ is 1 or $-1$ following that $\alpha $ is equal or different from
1 (see examples 2 and 3). These values correspond to the best value with
respect to the causality property (strictly verified only when $0<\alpha <1)$%
. Consequently, when the propagation is free of dispersion (the wave
celerity is constant whatever the frequency), the only possibility is $%
\alpha =2$ (the Gaussian case). It seems that this situation appears for
propagation in free space, together for acoustics and for electromagnetic
waves, though physical models are very different.

2) In literature, the Kramers-Kr\"{o}nig relations were proved under the
causality condition. If the (integrable) impulse response $f\left( t\right) $
\ of a filter verifies%
\begin{equation*}
f\left( t\right) =0,t<0
\end{equation*}%
and if its Fourier transform (the complex gain)%
\begin{equation*}
F\left( \omega \right) =\int_{-\infty }^{\infty }f\left( t\right)
e^{-i\omega t}dt
\end{equation*}%
is such that $\left\vert F\left( \omega \right) \right\vert \rightarrow
_{\left\vert \omega \right\vert \rightarrow \infty }0$ faster than $%
1/\left\vert \omega \right\vert ,$ the real and imaginary parts of $F\left(
\omega \right) $ are Hilbert transforms the one with respect to the other
(except for a sign). The proof is based on the fact that $F\left( z\right) $
is analytic in the lower half plane. Equivalently, the Laplace transform $%
F\left( z\right) $ of $f\left( t\right) $ exists for $y<0$ ($z=x+iy).$ In
this paper, the characteristic function $\psi \left( \omega \right) $ is the
complex gain. We know that, for the admitted values of $\beta $ (1 or $-1$
following the $\alpha $ value$),$ $\psi \left( z\right) $ has the property
of analicity and its behavior at $\pm \infty $ is fast enough \cite{Zolo}.
Consequently, the Kramers-Kr\"{o}nig relations are true, even if the
probability density of $\psi \left( \omega \right) $ does not cancel, which
is the case for $1\leq \alpha <2.$ In this case, the causality condition is
not verified.

3) We have admitted in many cases that $A\left( t\right) $ follows a stable
law with parameters $\left( l/c,al,\alpha ,\pm 1\right) .$ $l/c$ is a
parameter of position and $al$ is a scale parameter. The dependency in $l$
is a consequence of the Beer-Lambert law. When $1<\alpha \leq 2,$ the filter
linked to $\psi \left( \omega \right) $ is not causal, but this drawback
becomes more and more weak when $l$ increases, because $A\left( t\right) /l$
follows a stable law with parameters%
\begin{equation*}
\left( 1/c,al^{1-\alpha },\alpha ,\pm 1\right)
\end{equation*}%
which shows that the ratio scale parameter versus parameter of position is
more and more weak. For $\alpha =1,$ the ratio is constant.

\section{Conclusion}

I had shown for a long time and in this journal that random clock changes
provide interesting models for wave propagation \cite{Laca1}, \cite{Laca15}, 
\cite{Laca16}. Gaussian processes soon gave an interesting panel of
applications, but they often are not available outside the atmosphere.

The examples given in section 5 show that random propagation times following
Gaussian or non-Gaussian stable probability laws are good models in
acoustics, ultrasonics and in the domain of electromagnetic waves. Except
for the Gaussian, the used laws are not symmetric and the most possible
asymmetric (symmetric stable laws are referred as S$\alpha $S in literature
and widely used \cite{Samo}). The extreme case of asymmetry appears for the
values $\beta =\pm 1.$ This occurence is found in the given examples and it
is a strong argument for the model validity, because of links with the
causality of filters. The random propagation time \textbf{A }is not
sufficiently defined by its one-dimensional probability laws, and we need
insights about two-dimensional laws. The difficulty lies in the fact that
the order two moment of the r.v $A\left( t\right) $ does not exist, so that
the problem is outside the familiar framework of stationary processes with
finite autocorrelations and power spectra. In the appendix, we propose a
solution to this drawback.

\section{Appendix}

1) Let assume that the r.v. (random variables) $X_{n},n\in \mathbb{Z}$, are
(mutually) independent obeing same probability stable laws with parameters $%
m=0$ and any $a,\alpha ,\beta $ defined by $\left( 15\right) .$ Except when $%
\alpha =2$ (the Gaussian), E$\left[ X_{n}^{2}\right] =\infty .$ It is a
drawback which forbids the use of mean-square tools. We define the r.v. $Y$
from the sequence of positive real numbers \textbf{a=}$\left\{ a_{k},k\in 
\mathbb{Z}\right\} $ by%
\begin{equation*}
Y=\sum_{k=-\infty }^{\infty }a_{k}X_{k},a_{k}\geq 0.
\end{equation*}%
From the independence of the $X_{k}$ we have 
\begin{equation*}
\begin{array}{c}
\text{E}\left[ \exp \left( -\sum_{k=m}^{n}i\omega a_{k}X_{k}\right) \right] =
\\ 
\prod\limits_{k=m}^{n}\text{E}\left[ \exp \left( -i\omega a_{k}X_{k}\right) %
\right] = \\ 
\exp \left[ -a\left\vert \omega \right\vert ^{\alpha }c\left( \omega \right)
\sum_{k=m}^{n}a_{k}^{\alpha }\right] \text{ }%
\end{array}%
\end{equation*}%
with $c\left( \omega \right) =1-ib\tan \frac{\pi \alpha }{2}$sgn$\omega$ (for $\alpha \neq 1$).
The Levy continuity theorem learns us that the r.v $Y$ exists in the sense
of the convergence in law if and only if \cite{Luka} 
\begin{equation*}
\sum_{k=-\infty }^{\infty }a_{k}^{\alpha }<\infty .
\end{equation*}%
In this situation the probability law of $Y$ is stable with parameters
\begin{equation*}
\left( 0,a\sum_{k=-\infty }^{\infty }a_{k}^{\alpha },\alpha ,\beta \right) .
\end{equation*}%
When $\left\vert x\right\vert \rightarrow \infty ,$ the probability density
of the $X_{k}$ converges to 0 at least like $\left\vert x\right\vert
^{-1-\alpha }$ \cite{Luka}$.$ This allows to assert the existence a.s
(almost sure) of $Y$ by application of the three-series theorem of
Kolmogorov \cite{Loev}.

2) For any $h>0,$ we define \textbf{U}$_{h}=\left\{ U_{h}\left( t\right)
,t\in \mathbb{R}\right\} $ by%
\begin{equation}
U_{h}\left( t\right) =h^{1/\alpha }\sum_{k=-\infty }^{\infty }f\left(
t-kh\right) X_{k}
\end{equation}%
where $f\left( t\right) $ is some positive, even and regular enough
function, decreasing on $\mathbb{R}^{+}$. $U_{h}\left( t\right) $ is well
defined provided that 
\begin{equation*}
\sum_{k=-\infty }^{\infty }f^{\alpha }\left( t-kh\right) <\infty .
\end{equation*}%
The parameters of the stable law linked to $U_{h}\left( t\right) $ are%
\begin{equation*}
\left( 0,\theta _{h},\alpha ,\beta \right)
\end{equation*}%
where $\theta _{h}\left( t\right) $ is defined by%
\begin{equation*}
\theta _{h}\left( t\right) =ah\sum_{k=-\infty }^{\infty }f^{\alpha }\left(
t-kh\right)
\end{equation*}%
which can be chosen arbitrary close to the constant%
\begin{equation}
\theta _{0}=2a\int_{0}^{\infty }f^{\alpha }\left( u\right) du
\end{equation}%
when $h$ is small enough and assuming the existence of the integral. In this
case, $U_{h}\left( t\right) $ converges in law when $h\rightarrow 0$ to the
stable law%
\begin{equation*}
\left( 0,\theta _{0},\alpha ,\beta \right) .
\end{equation*}%
It is easy to construct simple examples which show that $U_{h}\left(
t\right) $ does not converge a.s when $h\rightarrow 0.$

3) For $t=2mh,\tau =2nh>0,$ we can write ($f$ is an even function)%
\begin{equation*}
\begin{array}{c}
U_{h}\left( t\right) -U_{h}\left( t-\tau \right) =h^{1/\alpha }\left[ A-B%
\right] \\ 
A= \\ 
\sum_{l=0}^{\infty }\left[ f\left( \left( n-l\right) h\right) -f\left(
\left( n+l\right) h\right) \right] X_{l+2m-n} \\ 
B= \\ 
\sum_{l=0}^{\infty }\left[ f\left( \left( n-l\right) h\right) -f\left(
\left( n+l\right) h\right) \right] X_{-l+2m-n}%
\end{array}%
\end{equation*}%
which leads to (the coefficients of the $X_{k}$ in $A$ and $B$ are positive
because $f\left( t\right) $ is assumed decreasing on $\mathbb{R}^{+})$%
\begin{equation*}
\left\{ 
\begin{array}{c}
\text{E}\left[ e^{-i\omega \left[ U_{h}\left( t\right) -U_{h}\left( t-\tau
\right) \right] }\right] = \\ 
\exp \left[ -2a\mu _{h}\left( \tau \right) \left\vert \omega \right\vert
^{\alpha }\right] \\ 
\mu _{h}\left( \tau \right) = \\ 
\sum_{l=0}^{\infty }\left[ f\left( \frac{\tau }{2}-lh\right) -f\left( \frac{%
\tau }{2}+lh\right) \right] ^{\alpha }%
\end{array}%
\right.
\end{equation*}%
$\mu _{h}\left( \tau \right) $ is independent of $t.$ When $h\rightarrow 0,$
we obtain%
\begin{equation}
\left\{ 
\begin{array}{c}
\lim_{h\rightarrow 0}\text{E}\left[ e^{-i\omega \left[ U_{h}\left( t\right)
-U_{h}\left( t-\tau \right) \right] }\right] = \\ 
\exp \left[ -2a\mu _{0}\left( \tau \right) \left\vert \omega \right\vert
^{\alpha }\right] \\ 
\mu _{0}\left( \tau \right) = \\ 
\int_{0}^{\infty }\left[ f\left( \frac{\tau }{2}-u\right) -f\left( \frac{%
\tau }{2}+u\right) \right] ^{\alpha }du.%
\end{array}%
\right.
\end{equation}%
Let note that the last probability law does not depend on $t.$

When $\alpha =2$ (the Gaussian case), we obtain a wide class of
characteristic functions but not the whole possibilities (Khinchine's
criterion, see \cite{Luka}). With this construction, it is possible to
obtain a behavior nearby the origin point like%
\begin{equation}
\left\{ 
\begin{array}{c}
\mu _{0}\left( \tau \right) =\left\vert \frac{\tau }{\tau _{0}}\right\vert
^{\gamma }+o\left( \left\vert \tau \right\vert ^{\gamma }\right) \\ 
\gamma =1\text{ when }0<\alpha <1 \\ 
1<\gamma <\alpha \text{ when }1<\alpha <2.%
\end{array}%
\right.
\end{equation}

\end{document}